\documentclass[amsmath,amssymb,
showpacs,twocolumn,
superscriptaddress,
prl]{revtex4-1}
\usepackage{graphicx,bm,color,subfigure}
\usepackage[T1]{fontenc}
\usepackage{mathrsfs}
\usepackage{bm}
\usepackage{amsmath}
\usepackage{amssymb}
\usepackage{graphicx}
\usepackage{esint}
\usepackage{multirow}
\usepackage{float}
\usepackage{array}
\usepackage{makecell}
\usepackage{harpoon}
\usepackage{booktabs}
\usepackage{gensymb}
\usepackage{simplewick}
\usepackage{subfigure}
\usepackage{soul}
\usepackage{makecell}

\begin{document}

\title{Nematic Superconductivity and Its Critical Vestigial Phases in the Quasi-crystal}
\author{Yu-Bo Liu}
\thanks{These two authors contributed equally to this work.}
\affiliation{School of Physics, Beijing Institute of Technology, Beijing 100081, China}
\author{Jing Zhou}
\thanks{These two authors contributed equally to this work.}
\affiliation{Department of Science, Chongqing University of Posts and Telecommunications, Chongqing 400065, China}
\affiliation{Institute for Advanced Sciences, Chongqing University of Posts and Telecommunications, Chongqing, 400065, China}
\author{Fan Yang}
\email{yangfan_blg@bit.edu.cn}
\affiliation{School of Physics, Beijing Institute of Technology, Beijing 100081, China}

\begin{abstract}
We propose a general mechanism to realize nematic superconductivity (SC) and reveal its exotic vestigial phases in the quasi-crystal (QC). Starting from a Penrose Hubbard model, our microscopic studies suggest that the Kohn-Luttinger mechanism driven SC in the QC is usually gapless due to violation of Anderson's theorem, rendering that both chiral and nematic SCs are common. The nematic SC in the QC can support novel vestigial phases driven by pairing phase fluctuations above its $T_c$. Our combined renormalization group and Monte-Carlo studies provide a phase diagram in which, besides the conventional charge-4e SC, two critical vestigial phases emerge, i.e. the quasi-nematic (Q-N) SC and Q-N metal. In the two Q-N phases, the discrete lattice rotation symmetry is counter-intuitively ``quasi-broken'' with power-law decaying orientation correlation. They separate the phase diagram into various phases connected via Berezinskii-Kosterlitz-Thouless (BKT) transitions. These remarkable critical vestigial phases, which resemble the intermediate BKT phase in the $q$-state ($q\ge 5$) clock model, are consequence of the five- (or higher-) fold anisotropy field brought about by the unique QC symmetry, which are absent in conventional crystalline materials.
\end{abstract}

\maketitle

{\bf Introduction:} The electron states in the quasicrystal (QC) are attracting more and more attentions recently~\cite{Bandres,Giergiel,Roberts,SSakai0,Hauck,SSakai1,Inayoshi,SSakai2,SSakai3,Lesser,Keskiner,Nagai,SSakai4,Jagennathan,Jagannathan1,Ciardi,Keskiner2}. Due to its special long-range order without translation period, the QC can host such as five- or eight-fold rotation symmetry forbidden in crystals. Various correlated~\cite{Wessel2003QC-QAF,Shaginyan2015QC-QPT,Thiem2015QC-QO,Andrade2015QC-QB,Otsuki2016QC-QCB, Koga2017QC-AFO,Miyazaki,Koga,Koga1,Ghosh,Sugimoto,Araujo} and topological~\cite{Kraus2012QC-TS, Huang2018QC-QSHS, Longhi2019QC-TPT,TPeng,Jeon,CWang,Bhola,Ghadimi2,RChen,YBYang} electron states have been revealed in the QC. Particularly, the discovery of superconductivity (SC) in the Al-Zn-Mg QC~\cite{Kamiya2018qSC} has aroused many interests recently~\cite{Sakai2017QC-SC, Hou2018Superfluid, Autti2018Superfluid, Araujo2019QC-SC, Sakai2019pairing, Takemori2020, Nagai2020BdG,GRai,Shiino,Khosravian,YBLiu,Fukushima,Fukushima1,YBLiu1}. Theoretically, the pairing symmetries in such QC as the 2D Penrose lattice have been classified~\cite{cao} according to the irreducible representation (IRRP) of the D$_5$ point group.  Remarkably, the 2D IRRPs can lead to chiral SC hosting spontaneous bulk current, driven by repulsive interaction via the Kohn-Luttinger (K-L) mechanism. Here we propose that gapless nematic SC can also be a common pairing phase in QCs. More interesting, partial melting of this order can lead to two critical vestigial phases, i.e. the quasi-nematic (Q-N) SC and Q-N metal, which are protected by the unique QC symmetry absent in crystals.

Generally in a pairing state belonging to the 2D IRRP of the point group, the two basis gap functions can be $1: i$ or $1:r$ ($r\in R$) mixed. In crystals, the $1: i$ mixing is usually energetically favored as it generates a full pairing gap~\cite{MengChen2010, Chubukov2012, HongYao2015}. However, the situation is distinct in QCs: It has been shown that, the Anderson's theorem~\cite{Anderson}, which states that an electron state tends to pair with its time-reversal partner, is violated in a K-L mechanism driven pairing phase in QCs~\cite{cao}. Here we show that the violation of this theorem usually leads to gapless SC, rendering that both chiral and nemtaic SCs are common in QCs, and we further focus on the finite-temperature vestigial phases~\cite{Agterberg,Berg,Agterberg1,Babaev2004,WHKo,Herland,FFSong,PLi,LFZhang,SZhou,Rampp,YYu,Curtis,Poduval,MZeng,MHecker,MHecker1,Jian2021,Liang_Fu,Grinenko2021,Wuyiming2023} of the nematic SC.

The nematic SC~\cite{JLi,Yonezawa,RTao,Kostylev,Chichinadze,TLe} spontaneously breaks the U(1)-gauge and lattice rotation symmetries. For the continuous U(1)-gauge symmetry, there exists a  Berezinskii-Kosterlitz-Thouless (BKT) transition temperature $T_{\text{BKT}}$ below which the pairing correlation power-law decays. For the discrete lattice-rotation symmetry, there usually exists a second-order transition temperature $T_{\text{nem}}$ below which long-range nematic order developes. When $T_{\text{BKT}}\ne T_{\text{nem}}$, two vestigial phases can emerge above $T_c$ of the nematic SC, i.e. the charge-4e SC or the nematic metal~\cite{Jian2021, Liang_Fu}. Here we demonstrate that for the nematic SC on the Penrose lattice, there exists an intermediate-temperature regime, wherein the discrete lattice-rotation symmetry is counter-intuitively ``quasi-broken'', leading to extended critical vestigial phases with power-law decaying orientation correlations, dubbed as Q-N phases.

In this paper, we start from a Penrose Hubbard model. Based on the K-L mechanism, our microscopic calculations suggest that the violation of Anderson's theorem usually leads to gapless SC with finite zero-energy density of state (DOS). For the 2D IRRPs of D$_5$, our combined Ginzburg-Landau (G-L) analysis and microscopic energy calculations can lead to either chiral or nematic SCs for different parameters. We then study the vestigial phases of the nematic SC driven by the phase fluctuations of the two pairing components, via combined renormalization group (RG) and Monte-Carlo (MC) approaches. In the obtained phase diagram, besides the charge-4e SC, two critical vestigial phases emerge, i.e. the Q-N SC and Q-N metal (MT), which render that all phase transitions are BKT like. The two remarkable critical phases, which resemble the intermediate BKT phase of the $q$-state ($q\ge 5$) clock model, are brought about by the five- (or higher-) fold anisotropy field caused by the unique QC symmetry, which are absent in crystals.

\begin{figure}[htbp]
\includegraphics[width=0.45\textwidth]{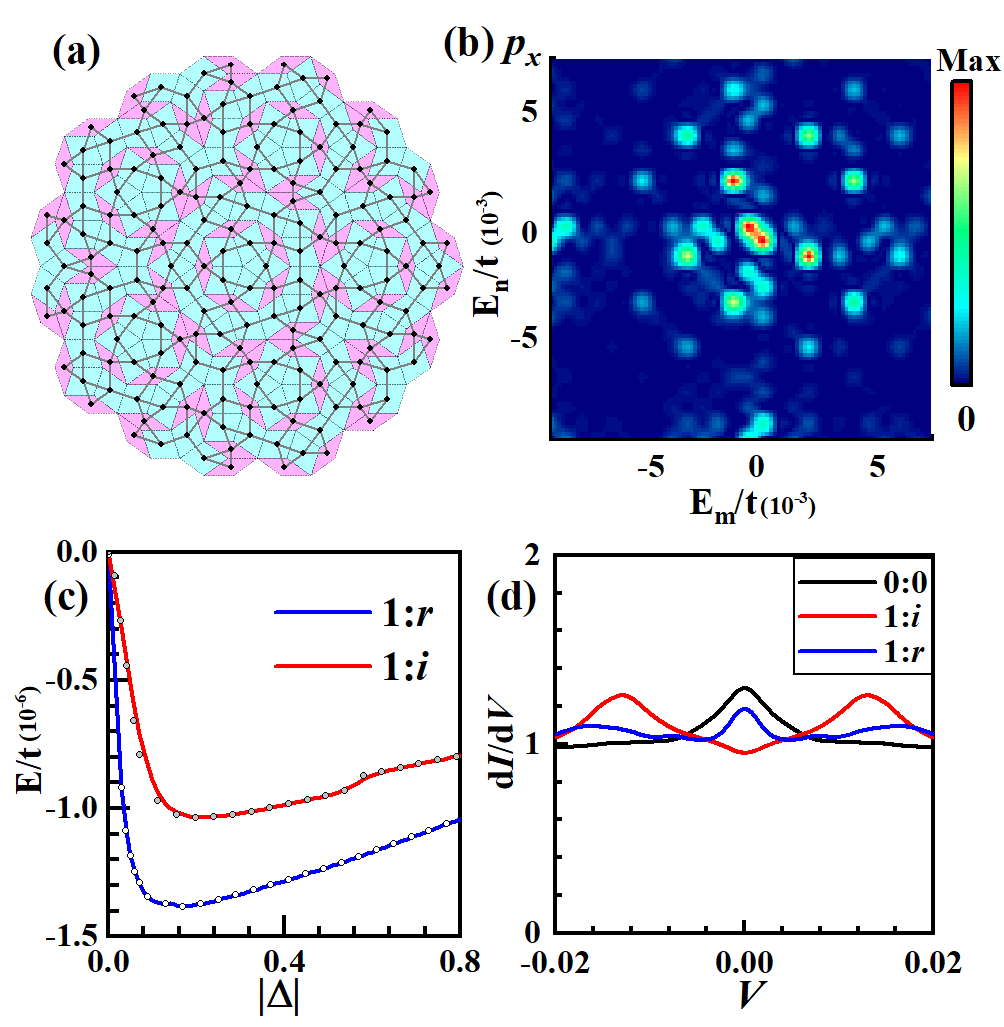}
\caption{(Color online) (a) Schematic diagram of Penrose lattice. The lattice sites at the center of the rhombuses are marked by solid black circles. NN bonding is marked by solid gray lines. (b) Contour plots of relative $|\Delta_{mn}|$, for a singlet $p_x$-wave pairing for $\delta=0.49$ and $U/W_D = 0.5$. (c) Ground state energy $E$ as function of the global magnitude $|\Delta|$ for the pairing with $1:r$ (bule, minimized for $r$) and $1: i$ (red) mixing of the two degenerate basis functions. The energy $E$ is in unit of $t$ in (b-c). (d) STM dI/dV of a typical site for the nematic SC (bule), chiral SC(red) and normal state(black). %(e) the specific heat $Cv$ for the nematic SC as function of $T$. (f) The current $J$ as a function of vector potential $\mathbf{A}$ at several temperatures.
}\label{microscopic}
\end{figure}

{\bf Model and Gapless Nematic SC:} Let us consider the following Hubbard model on the Penrose lattice,
\begin{eqnarray}\label{H0}
H=-\sum_{<i,j>\sigma}tc^{\dagger}_{i\sigma}c_{j\sigma}+U\sum_{i}n_{i\uparrow}n_{i\downarrow}-\mu\sum_{i,\sigma}n_{i\sigma},
\end{eqnarray}
where $c_{i\sigma}$ annihilates an electron at site $i$ with spin $\sigma$, $n_{i\sigma}$ is the electron-number operator, and $\mu$ denotes the chemical potential. Here the lattice sites are defined as the centers of the rhombuses on the Penrose tiling, as marked by the black solid circles in Fig.~\ref{microscopic}(a). We define two rhombuses sharing an edge as nearest neighbor (NN) \cite{tsunetsugu,tsunetsugu2}, and only consider hoppings along the NN bonds, as marked by the solid lines in Fig.~\ref{microscopic}(a).  The tight-binding  part of Eq. (\ref{H0}) is diagonalized as $H_{\text{TB}}=\sum_{m\sigma}\tilde{\epsilon}_{m} c^{\dagger}_{m\sigma}c_{m\sigma}$, with $c_{m\sigma}=\sum_i\xi_{i,m}c_{i\sigma}$. Here $m$ labels a single-particle eigen state with eigen-energy $\tilde{\epsilon}_{m}=\epsilon_{m}-\mu$ and eigenstate $\xi_{i,m}$. The total band width is $W_D\approx 8t$. We adopt $U=0.5W_D$ in our calculations.

The Cooper pairing in this system can be driven by the K-L mechanism~\cite{kohn,Baranov}, generalized to the cases in the QC~\cite{cao}. In this mechanism, two electrons near the Fermi level can gain effective attraction through exchanging particle-hole excitations in several second-order perturbative processes. Then a BCS mean-field (MF) treatment on the obtained effective Hamiltonian provides the self-consistent gap equation, which after linearized near $T_c$ takes the form of an eigenvalue problem of the interaction matrix. The $T_c$ is given by the temperature at which the largest eigenvalue of this matrix attains one, and the pairing symmetry, classified according to the IRRPs of D$_5$, is determined by the corresponding eigenvector. See the Supplementary Materials (SM)~\cite{SM} for details.

In Fig.~\ref{microscopic}(b), we show distribution of the amplitude $|\Delta_{mn}|$ ($\Delta_{mn}\in R$ ) of a typical singlet $p_x$-wave pairing gap function between the states $m$ and $n$ (labeled by their energies) near the Fermi level, obtained at the filling $\delta=0.49$. That of the $p_y$- symmetry in the same 2D $(p_x,p_y)$ IRRP is given in the SM~\cite{SM}. Fig.~\ref{microscopic}(b) displays that for each $m$, there is no unique $n$ rendering $|\Delta_{mn}|$ dominates that of any other $n$, violating Anderson's theorem. The BCS-MF Hamiltonian for this pairing state reads
\begin{eqnarray}\label{BCS_MF}
H_{\text{BCS-MF}}&=&\sum_{m\sigma}\tilde{\epsilon}_{m}c^{\dagger}_{m\sigma}c_{m\sigma}\nonumber\\&+&\sum_{m,n}\left(c_{m\uparrow}^{\dagger}c_{n\downarrow}^{\dagger}-
c_{m\downarrow}^{\dagger}c_{n\uparrow}^{\dagger}\right)\Delta_{mn}+h.c.
\end{eqnarray}
If $\Delta_{mn}=\Delta_{m}\delta_{mn}$, Eq. (\ref{BCS_MF}) is diagonalized to yield the Bogoliubov quasi-particle dispersion $E_m=\sqrt{\tilde{\epsilon}_{m}^2+\Delta_{m}^2}$, under which the condition $E_m=0$ leads to two combined equations: $\tilde{\epsilon}_{m}=0; \Delta_{m}=0$. In 2D at thermal-dynamic limit, the two equations lead to at most isolate solutions for $m$, corresponding to point gap nodes or full gap. However, due to violation of Anderson's theorem here, $E_m$ no longer takes this simple analytical form. Consequently, $E_m=0$ only provides one equation, which in 2D usually leads to an $O(L)$ ($L$: lattice size) number of $m$, forming a gapless SC carrying finite zero-energy DOS.

The mixing ratio between the two basis gap functions of a 2D IRRP, e.g. $(\Delta_{p_x},\Delta_{p_y})$, is analyzed via the G-L theory given in the SM~\cite{SM}. For convenience, we rotate the bases as $\Delta_{\pm}=\Delta_{p_x}\pm i\Delta_{p_y}$. The transformation of $\Delta_{\pm}$ under the $C_5^1$ rotation is $\hat{P}_{\frac{2\pi}{5}}\Delta_{\pm}(\mathbf{r})=e^{\pm 2i\pi /5}\Delta_{\pm}(\mathbf{r})$. Under the mirror reflection, $\Delta_{\pm}$ mutually exchange.  The mixed gap function is $\Delta=\psi_{+}\Delta_{+}+\psi_{-}\Delta_{-}$. Fixing $\Delta_{\pm}$, the G-L free energy $F=F(\psi_{+},\psi_{-})$ can only take the following $D_5\otimes U(1)$-gauge symmetry-allowed form~\cite{SM}
\begin{eqnarray}\label{G_L_F}
F(\psi_{+},\psi_{-})&=&\alpha(|\psi_{+}|^2+|\psi_{-}|^2)+\beta(|\psi_{+}|^4+|\psi_{-}|^4)\nonumber\\&&+\gamma|\psi_{+}|^2|\psi_{-}|^2+O(\psi_{\pm}^6)
\end{eqnarray}
If $\gamma>2\beta$, $F$ is minimized at $\psi_{+}=0$ or $\psi_{-}=0$, leading to a chiral SC wherein $\Delta_{p_x}$ and $\Delta_{p_y}$ are $1:i$ mixed; if $\gamma<2\beta$, $F$ is minimized at $|\psi_{+}|/|\psi_{-}|=1$, leading to a nematic SC wherein $\Delta_{p_x}$ and $\Delta_{p_y}$ are $1:r$ mixed ($r\in R$).

To determine the realized ground state, we calculate the energy $E$ as function of the global amplitude $\Delta$ for the $1:r$ (minimized for $r$) and $1:i$ mixing cases. As shown in Fig.~\ref{microscopic}(c), the energy of the $1:r$ mixing is lower, suggesting a nematic SC ground state. This result seems conflicting with the intuition that the chiral SC is usually energetically favored due to opening of a full pairing gap~\cite{MengChen2010}. This counter-intuitive result can be explained by Fig.~\ref{microscopic}(d) which displays the local DOS detected by the STM $dI/dV$ curve for a typical site (that for more sites are given in the SM~\cite{SM}). Fig.~\ref{microscopic}(d) shows that both the chiral and nematic SCs are gapless. Therefore in QCs, the chiral SC loses its advantage in energy, rendering that the nematic SC is also common. Note that chiral SC is also possible in this system, see the case at $\delta=0.51$~\cite{SM}. The gapless SC resembles the standard Fermi liquid in nature of elementary excitations, reflected in such quantities as the linearly temperature-dependent specific heat and saturate Knight-shift when $T\to 0$. However, this state carries nonzero superfluid density. See the SM~\cite{SM}.

{\bf Phase Diagram and Vestigial Phases:} Above the $T_c$ of nematic SC, nontrivial vestigial phases can be driven by the phase fluctuations of its two pairing components~\cite{Jian2021, Liang_Fu}. Under thermal fluctuations, the global amplitudes $\psi_{\pm}$ appearing in Eq.~(\ref{G_L_F}) become functions of the coarse-grained position $\mathbf{r}$. Despite lack of translation period, the QC is uniform in the long-wave limit~\cite{CWang,KJiang}. Therefore, $\psi_{\pm}(\mathbf{r})$ is smooth function of $\mathbf{r}$.  Focusing on low-energy phase fluctuations, we set $\psi_\pm(\mathbf{r})=\psi_{0}e^{i\theta_{\pm}(\mathbf{r})}$, with the constant $\psi_{0}>0$ and pairing phases $\theta_{\pm}(\mathbf{r})\in (0,2\pi)$. To include dependence on $\theta_{\pm}(\mathbf{r})$, the free energy functional $F$ is expanded to $O(\psi_\pm^{10})$ as~\cite{SM}
\begin{eqnarray}\label{G_L_F_10}
F^{(10)}\left(\psi_{+},\psi_{-}\right)=-A_{0}\left(\psi_{+}^5\psi_{-}^{5*}+{\rm c.c.}\right)+O\left(\psi^{12}\right).
\end{eqnarray}
Let's introduce the global and relative phase fields $\theta$ and $\phi$ through $\theta_{\pm}\left(\mathbf{r}\right)=\theta\left(\mathbf{r}\right)\pm\phi\left(\mathbf{r}\right)$. Physically, ordering of the $\theta$ field breaks the U(1)-gauge symmetry and represents for SC, while ordering of the $\phi$ field breaks the rotation symmetry and indicates the orientation (nematic) order. When dependence on $\triangledown\theta$ and $\triangledown\phi$ is included~\cite{Jian2021}, the low-energy classical Hamiltonian is given as,
\begin{eqnarray}\label{Hamiltonian_r}
H=\int d^{2}\mathbf{r}\left(\frac{\rho}{2}\left|\triangledown\theta\right|^2+ \frac{\kappa}{2}\left|\triangledown\phi\right|^2-A\cos 10\phi\right).~~
\end{eqnarray}
Here $\rho/\kappa$ are stiffness parameters, and $A=2A_0\psi_0^{10}$.

Eq.~(\ref{Hamiltonian_r}) shows that, while the Hamiltonian for the $\theta$ field describes a continuous-space pure XY model, that for the $\phi$ field describes a continuous-space XY model subject to a $q$-fold ($q=5$) anisotropy field, resembling the $q$-state clock model in symmetry. Note that $(\theta\left(\mathbf{r}\right),\phi\left(\mathbf{r}\right)+\pi)$ and $(\theta\left(\mathbf{r}\right),\phi\left(\mathbf{r}\right))$ describe gauge equivalent states as their corresponding physical $(\theta_{+}\left(\mathbf{r}\right), \theta_{-}\left(\mathbf{r}\right))$ configurations are only globally different by a constant $\pi$~\cite{Yu_Bo_Liu2023}. Therefore, the seeming ten saddle points for the $\phi$ field in Eq.~(\ref{Hamiltonian_r}) actually represent for five ones, causing the five-fold anisotropy. In (\ref{Hamiltonian_r}), the $\theta$ and $\phi$ fields are subject to the constraint that both fields should host integer or half-integer vortices simultaneously~\cite{Agterberg,Berg, Agterberg1, Babaev2004}.

\begin{table}[!h]
\label{tab:1}
\centering
\caption{Correspondence between RG fixed points and phases. The new abbreviations denote: MT (normal metal), 4e-SC(charge-4e SC), N-SC (nematic SC).}\label{tab:1}
\begin{tabular}{|c|c|c|c|c|c|}
  \hline\hline
  $g_{2,0}$ & $\infty$ & 0 & $\infty$ & 0 & 0 \\
  \hline
  $g_{0,2}$ & $\infty$ & $\infty$ & 0 & 0 & 0 \\
  \hline
  $g_{1,1}$ & $\infty$ & 0 & 0 & 0 & 0 \\
  \hline
  $g_{10}$ & 0 & 0 & 0 & 0 & $\infty$ \\
  \hline
  %phase & normal & normal & normal & charge-4e SC & quasi-nematic metal & quasi-nematic SC & nematic SC \\
  phase & MT & 4e-SC & Q-N MT & Q-N SC & N-SC \\
  \hline\hline
\end{tabular}
\end{table}

We employ the RG approach to study the model (\ref{Hamiltonian_r}), and map it to a dual two-component Sine-Gordon model described by the following action~\cite{Jian2021},
\begin{eqnarray}
\label{eqn:action-SG}
S_{\mathrm{SG}}&&=\int d^{2}\mathbf{x}\left( \frac{T}{2\rho}\left|\triangledown\tilde{\theta}\right|^2+\frac{T}{2\kappa}\left|\triangledown\tilde{\phi}\right|^2-g_{10}\cos10\phi -g_{2,0}\right.\nonumber\\
&&\left.\times\cos2\pi\tilde{\theta}-g_{0,2}\cos2\pi\tilde{\phi}-g_{1,1}\cos\pi\tilde{\theta}\cos\pi\tilde{\phi}\right),
\end{eqnarray}
where $\tilde{\theta}/\tilde{\phi}$ are dual vortex fields of $\theta/\phi$. Here $g_{2,0}/g_{0,2}$ are fugacities for integer $\tilde{\theta}/\tilde{\phi}$ vortices while $g_{1,1}$ is that for half $\phi$-half $\theta$ vortices, and $g_{10} \propto A$ is the 5-fold anisotropy parameter. While details of the RG approach including the one-loop RG flow equation are provided in the SM~\cite{SM}, the correspondence between the available fixed points and the phases are listed in Tab. \ref{tab:1}.

\begin{figure}[h]
	\centering
	\includegraphics[width=0.45\textwidth]{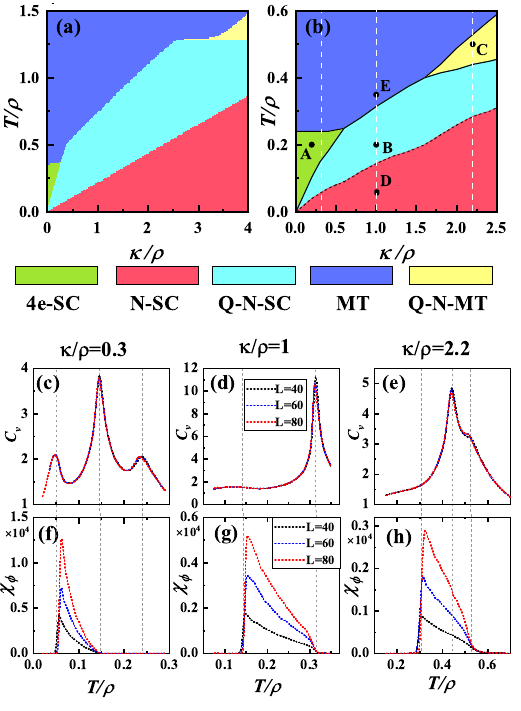}
	\caption{(Color online) Phase diagrams provided by (a) the RG- and (b) the MC- studies. The initial values of the coupling parameters for obtaining (a) are $g_{2,0}=g_{0,2}=0.1$, $g_{1,1}=0.01$, $g_{10}=0.001$ and for (b) are $A=0.025\rho$. The white dashed lines in (b) represent $\kappa/\rho=0.3,~1,~2.2$, respectively. (c-e) Specific heat $C_v$ and (f-h) the $\phi$-field susceptibility $\chi_{\phi}$ as function of temperature $T/\rho$ on different lattice sizes ($L=40,~60,~80$) for $\kappa/\rho=0.3$ in (c) and (f), $\kappa/\rho=1$ in (d) and (g), and $\kappa/\rho=2.2$ in (e) and (h). The grey dashed lines in (c)-(h) mark the phase transitions.}\label{phase_diagram}
\end{figure}

The RG phase diagram is shown in Fig.~\ref{phase_diagram}(a), which is topologically insensitive to the initial values of the coupling parameters~\cite{SM}. When $T\to 0$, all fugacities are irrelevant while $g_{10}$ is relevant, forming the nematic SC (N-SC). When $T$ arises, the system first enters the Q-N SC when $g_{10}$ becomes irrelevant. When $T$ further enhances, if $\kappa<<\rho$, the Q-N SC  turns into the charge-4e SC (4e-SC) once $g_{0,2}$ gets relevant rendering proliferation of the $\phi$ vortices; if $\kappa>>\rho$, the Q-N SC turns into the Q-N MT once $g_{2,0}$ gets relevant rendering proliferation of the $\theta$ vortices. When $T$ is high enough, the normal MT is achieved for whatever $\kappa/\rho$. If $\kappa\approx\rho$, when $T$ arises, the Q-N SC directly turns into the MT once $g_{1,1}$ gets relevant rendering proliferation of the half $\phi$-half $\theta$ vortices.

{\bf Quasi-Nematic Phases:} Two new phases absent in previous studies~\cite{Jian2021, Liang_Fu} emerge in the phase diagram Fig.~\ref{phase_diagram}(a) and Tab.~\ref{tab:1}: the Q-N SC and Q-N MT. These two Q-N phases are realized when the fugacity $g_{0,2}$ is irrelevant so that no free $\phi$-vortex is excited, but the anisotropy parameter $g_{10}$ for the $\phi$-field is irrelevant. To further study the nature of the two new phases and their phase transitions, we perform a MC study~\cite{SM} on a discretized version of the continuous Hamiltonian (\ref{Hamiltonian_r}). The obtained specific heat, superfluid density, susceptibilities, Binder cumulants and correlation functions~\cite{SM} combinedly provide the phase diagram shown in Fig.~\ref{phase_diagram}(b), which is topologically consistent with Fig.~\ref{phase_diagram}(a).

Taking three typical $\kappa/\rho=0.3,~1,~2.2$ marked in Fig.~\ref{phase_diagram}(b) , we display the temperature $T/\rho$ dependence of the specific heat $C_v$ and the $\phi$-field susceptibility $\chi_{\phi}$ on different lattice sizes ($L=40,~60,~80$) in Fig.~\ref{phase_diagram}(c-e) and (f-h), respectively. The grey dashed lines in (c-h) mark the phase transitions. For $C_v$ (c-e), the phase transitions either showcase as broad humps or are featureless, which are insensitive to $L$, implying that no singularity will emerge upon $L\to \infty$, suggesting that all transitions are BKT-like. While it's known that the 2D SC transition is BKT-like, here it's remarkable that the phase transitions related to the breaking of the discrete lattice-rotation symmetry are also BKT-like. This point is related to the $T/\rho$ dependence of $\chi_{\phi}$ (f-h): While it's finite and small in the low-$T$ nematic phase (N-SC) and high-$T$ non-nematic phases (4e-SC and MT), it strongly depends on $L$ and diverges upon $L\to \infty$ in the intermediate-$T$ Q-N phases (Q-N SC and Q-N MT) resembling the divergence of $\chi_\theta$ in the SC phases~\cite{SM}, suggesting that the Q-N phases are BKT-like extended critical phases for the $\phi$-field. The transitions from the Q-N phases to the nematic or non-nematic phases are BKT-like.

\begin{figure}[h]
	\centering
	\includegraphics[width=0.45\textwidth]{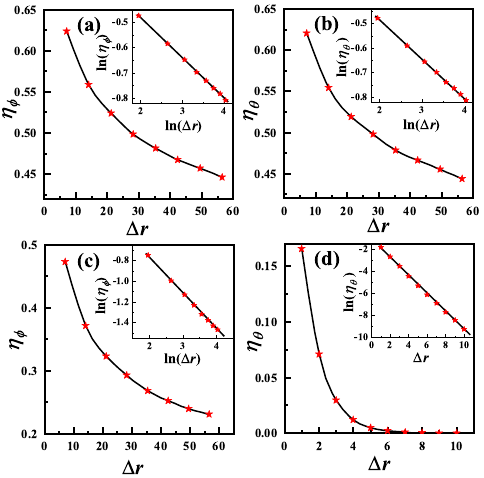}
	\caption{(Color online) The correlation function $\eta_{\phi/\theta}$ for (a) and (b) for the point $\mathbf{B}$ ($\kappa=0.1\rho, T=0.2\rho$), for (c) and (d) for  the point $\mathbf{C}$ ($\kappa=2.2\rho, T=0.5\rho$) marked in Fig.~\ref{phase_diagram}(b). Insets of (a-c) the log-log plot, and (d) only the y- axis is logarithmic. The $\eta_{\phi/\theta}$ for A, D, E are given in the SM~\cite{SM}.}\label{ABC}
\end{figure}

The nature of the two Q-N phases is reflected in the correlation functions $\eta_{\theta}(\Delta \mathbf{r})\equiv\frac{1}{N}\sum_{\mathbf{r}}\left\langle e^{i[\theta(\mathbf{r})-\theta(\mathbf{r}+\Delta \mathbf{r})]}\right\rangle$ and $\eta_{\phi}$ defined similarly. Fig.~\ref{ABC}(a) and (b) show $\Delta r$ ($\equiv |\Delta \mathbf{r}|$)-dependence of $\eta_{\phi}$ and $\eta_{\theta}$ for the typical point B marked in Fig.~\ref{phase_diagram}(b). Obviously, both $\eta_{\theta}$ and $\eta_{\phi}$ power-law decay with $\Delta r$, reflecting the Q-N SC. Fig.~\ref{ABC}(c) and (d) are for the typical point C marked in Fig.~\ref{phase_diagram}(b): While $\eta_{\theta}$ decays exponentially with $\Delta r$, $\eta_{\phi}$ power-law decays with $\Delta r$, reflecting the Q-N MT. The common feature for both Q-N phases is power-law decaying of the orientation correlation $\eta_{\phi}\sim \Delta r$, indicating the quasi-long-range order of the $\phi$ field, suggesting remarkable ``quasi-breaking'' of the discrete lattice-rotation symmetry.

{\bf Discussion and Conclusion:} The counterintuitive Q-N phases obtained here bear resemblance to the intermediate BKT phase in the 2D $q$-state clock model for $q\ge5$~\cite{Jose1977,Tobochnik1982,Challa1986,Surungan2019,Ziqian_Li2020,Hao_Chen2020,Miyajima2021}, which also exhibits power-law decaying correlation and BKT transitions to adjacent phases. Such intriguing phase fluctuation driven Q-N phases can only emerge on QCs: As derived in the SM~\cite{SM}, for a $D_{2n}$ ($D_{2n+1}$) symmetric lattice, the anisotropy-field Hamiltonian for the $\phi$ field is $-A\cos \left(2n\phi\right)$ ($-A\cos \left[2(2n+1)\phi\right]$), leading to the $n$ ($2n+1$) fold anisotropy, resembling the $n$ ($2n+1$)-state clock model in symmetry. Consequently, only the $D_5$, $D_7$ or $D_n (n\ge 9)$ lattices can host the Q-N phases, which can only be realized on QCs.

In conclusion, the SC driven by K-L mechanism in the QC violates Anderson's theorem, leading to possible nematic SC. Our combined RG and MC calculations reveal the emergence of novel Q-N vestigial phases protected by the unique QC symmetry. These vestigial phases are BKT-like extended critical phases with power-law decaying orientation correlation, resembling the intermediate BKT phase in the $q$-state ($q\ge 5$) clock model.

\section*{Note:}Shortly after this work was finished and announced, another highly related work emerged~\cite{Fernandes2024}, in which the ``critical nematic phase'' (which has the same physical meaning as the ``quasi-nematic phase'' dubbed here) is revealed as possible vestigial phase of the nematic SC on the 30\degree-twisted hexagonal bilayer, which hosts 12 fold quasi-crystal rotation symmetry.

\section*{Acknowledgements:} This work is supported by the NSFC under the Grant Nos. 12234016, 12074031 and the funding of Institute for Advanced Sciences of Chongqing University of Posts and Telecommunications (E011A2022326).

%as detected by the

\appendix
\begin{widetext}
\section{Microscopic Calculations Based on Kohn-Luttinger Mechanism}
The microscopic calculations start from the standard repulsive Hubbard model on the Penrose lattice. The Cooper pairing can be driven by the Kohn-Luttinger (K-L) mechanism~\cite{KL1,KL2}, generalized to the cases on the QC~\cite{cao}. In the K-L mechanism, two electrons near the Fermi level can gain effective attraction through exchanging particle-hole excitations in several second-order perturbative processes, leading to the following effective Hamiltonian:
\begin{eqnarray}\label{H0}
H_{eff}=-\sum_{<i,j>\sigma}tc^{\dagger}_{i\sigma}c_{j\sigma}+U\sum_{i}n_{i\uparrow}n_{i\downarrow}-\mu\sum_{i,\sigma}n_{i\sigma}-(U^2/2)\sum_{ij\sigma\sigma'}\chi_{ij}c^{\dagger}_{i\sigma}c_{i\sigma'}c^{\dagger}_{j\sigma'}c_{j\sigma},
\end{eqnarray}
where $c_{i\sigma}$ annihilates an electron at site $i$ with spin $\sigma$, $n_{i\sigma}$ is the electron-number operator, and $\mu$ denotes the chemical potential. $\chi_{ij}$ is the static susceptibility, defined as
\begin{eqnarray}\label{chi}
\chi_{ij}=\sum_{mn}\xi_{i,m}\xi_{j,m}\xi_{i,n}\xi_{j,n}\dfrac{n_F(\tilde{\epsilon}_m)-n_F(\tilde{\epsilon}_n)}{\tilde{\epsilon}_n-\tilde{\epsilon}_m}
\end{eqnarray}
Here $m$ labels a single-particle eigen state with eigen-energy $\tilde{\epsilon}_{m}=\epsilon_{m}-\mu$ and $\xi_{i,m}$ represents for the wave function of the state $m$. $n_F$ is the Fermi-Dirac function. A BCS mean-field (MF) study on the effective Hamiltonian leads to the self-consistent gap equation, which reduces to the following linearized equation at $T_{c}$,
\begin{eqnarray}\label{linearized}
\sum_{m'n'}F^{(s/t)}_{mn,m'n'}\tilde{\Delta}_{m'n'}=\tilde{\Delta}_{mn}
\end{eqnarray}
where $m,n,m',n'$ are the state indices and $s/t$ labels spin singlet/triplet state. See Ref~\cite{cao} for the details of the interaction matrix $F^{(s/t)}$. We just consider the Cooper pairing $\Delta_{mn}$ taking place near the Fermi surface while the $m,n$-states belong to a narrow energy shell near the Fermi level. $T_c$ is the temperature at which the largest eigenvalue of $F^{s/t}$ matrix attains one, and the pairing symmetry is determined by the corresponding eigenvector. The possible pairing symmetries can be classified according to the IRRPs of the D$_5$ point group, including 1D and 2D IRRPs. Note that the spin statistics and pairing symmetry are independent, i.e. each IRRP can have either spin-singlet or spin-triplet pairing. See Ref~\cite{cao} for more details.

In the rest of this section, we present more calculation results. In subsection A, we present the distribution of the $p_y$-wave and $s$-wave gap functions near the Fermi level in the state space. In subsection B, we present the results of some experimental quantities of the gapless SC obtained in our work, including the STM spectrum, the specific heat, the NMR Knight-shift and the superfluid density. In subsection C, we present the results of the case in which the gapless chiral SC as the ground state while filling  that the ground state for another filling $\delta=0.51$.

\subsection{Typical Gap Functions}

In Fig.~\ref{anderson}, as a supplement to the Fig.1(b) in the main text, we show distribution of the amplitude $|\Delta_{mn}|$ ($\Delta_{mn}\in R$ ) of a typical singlet $p_y$- and $s$-wave pairing gap functions between the states $m$ and $n$ (labeled by their energies) near the Fermi level, with the filling $\delta=0.49$. For each $m$, there is no unique $n$ which makes $|\Delta_{mn}|$ dominate that of any other $n$, violating Anderson's theorem.

\begin{figure}[h]
	\centering
	\includegraphics[width=0.8\textwidth]{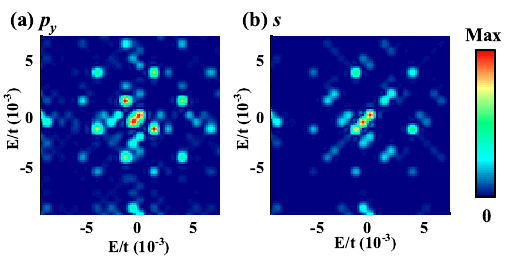}
	\caption{(Color online) Contour plots of relative $|\Delta_{mn}|$, for a singlet $p_y$-wave(a) and s-wave(b) state for $\delta=0.49$ and $U/W_D = 0.5$. The state $m,n$ are labeled by their energies $E$ in unit of $t$. }\label{anderson}
\end{figure}

\subsection{Experiment Quantities}
In order to investigate the superconducting properties in QCs, we write out the B-dG (Bogoliubov-de Gennes) Hamiltonian matrix in the state space,
\begin{eqnarray}
H_{\text{BCS-MF}} &=&
\sum_{m=1,\sigma}^{N_{c}}\tilde{\varepsilon}_{m}c^{\dagger}_{m\sigma}c_{m\sigma}+\sum_{mn=1}^{N_{c}}\left(c_{m\uparrow}^{\dagger}c_{n\downarrow}^{\dagger}-
c_{m\downarrow}^{\dagger}c_{n\uparrow}^{\dagger}\right)\Delta_{mn}+h.c.\nonumber    \\
~&=&
\left(\begin{array}{ccccc}
...& c_{m\uparrow}^{\dagger}&...& c_{m\downarrow}&...\end{array}\right)\left(\begin{array}{cc}
\tilde{\varepsilon} & \Delta\\
\text{\ensuremath{\Delta^{\dagger}}} & -\tilde{\varepsilon}
\end{array}\right)\left(\begin{array}{c}
...\\\hat{c}_{n\uparrow}\\
...\\
\hat{c}_{n\downarrow}^{\dagger}\\
...
\end{array}\right)\nonumber    \\
~&=&
X^{\dagger}H_{BdG}X=\sum_{l=1}^{2N_{c}}E_{l}\gamma_l^{\dagger}\gamma_l.
\end{eqnarray}
where $m,n$-states belong to a narrow energy shell near the Fermi level. Here the thickness of the energy shell is $0.06t$ and it includes $N_{c}(=100)$ states. In subsequent text, $m,n$ just represents states in the energy shell. The Bogoliubov transformation is written as $X=\Omega\gamma$. The amplitude of SC order parameter $|\Delta_{mn}|$ can be determined by the free energy minimization approach at finite temperatures. The expression of the free energy is

\begin{eqnarray}\label{F}
F=E-TS.
\end{eqnarray}
where the ground energy $E$ is the expectation value of the effective Hamiltonian, and the entropy $S=K_B\sum_{l}ln(1+e^{-\beta E_l})+\beta E_ln_f(E_s)$, where $\beta=1/K_BT$.

Fig.1(c) in the main text shows the SC ground state energy as a function of $|\Delta|$. It indicates that the ground state is the nematic SC when $\delta = 0.49$ and $U/W_D = 0.5$. After determining the global amplitude $|\Delta|$ of the SC order parameter by the free energy minimization approach, we investigate some experimental quantities of the nematic SC state for $\delta = 0.49$ and $U/W_D = 0.5$ , including the following

1) The scanning tunneling microscopy (STM) dI/dV spectrum at site $j$ can be written as
\begin{eqnarray}\label{stm}
D(\omega) = \int \sum_{\sigma}\left<T_\tau c_{j\sigma}^{\dagger}(\tau)c_{j\sigma}(0)\right>e^{i\omega\tau} d \tau
\end{eqnarray}
The STM dI/dV spectrum are site dependent, distinct from the periodic lattice. The STM dI/dV curve for a typical site is shown in the Fig. 1(d) in the main text. For generality, Fig.~\ref{stms} shows the STM dI/dV curve on additional typical sites for both nematic SC(bule line), chiral SC(red line) and normal state(black line), and all STM dI/dV curve in the main text and Fig.~\ref{stms} indicates that both the nematic and chiral SC states in this model can be gapless.

\begin{figure}[h]
	\centering
	\includegraphics[width=0.95\textwidth]{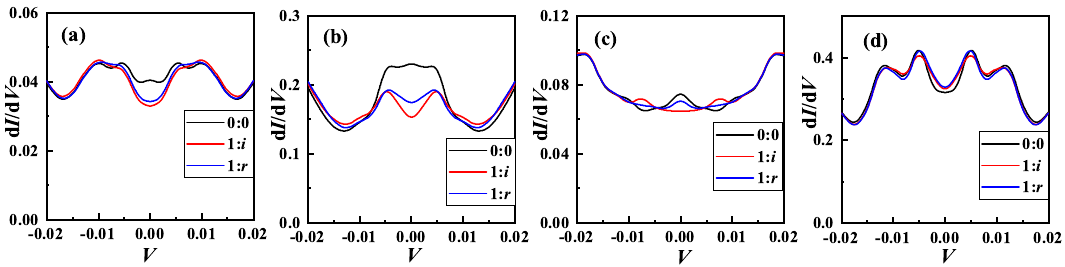}
	\caption{(Color online) The STM dI/dV spectra of some typical sites for the nematic SC(bule line), the chiral SC(red line) and the normal state(black line). $V$ is in unit of $t$.}\label{stms}
\end{figure}

2) The specific heat $C_v$ is given by
\begin{eqnarray}\label{Cv1}
C_v=T\dfrac{\partial S}{\partial T}
\end{eqnarray}

Fig.~\ref{cvjk}(a) shows the specific heat for the nematic SC as a function of temperature $T$. In the low-$T$ region, except for a tiny finite-size gap, the specific heat is proportional to temperature, similar to the behavior in Fermi liquid (FL).

3) The Knight-shift is given by
\begin{eqnarray}\label{K}
K=\int \left<T_\tau S^{+}(\tau)S^{-}(0)\right>e^{i\omega\tau} d \tau
\end{eqnarray}
where $S^+=\sum_{i}c_{i\uparrow}^{\dagger}c_{i\downarrow}$ and $S^-=\sum_{i}c_{i\downarrow}^{\dagger}c_{i\uparrow}$.  Fig.~\ref{cvjk}(b) exhibits that the NMR Knight-shift $K$ for nematic SC saturates to a finite value in the low temperature region, similarly to the Pauli-susceptibility behavior for standard FL.

4) The superfluid density $\rho$ is related to the current $J$ given by
\begin{eqnarray}\label{J}
J_{\alpha}(\mathbf{A})=-\sum_{<ij>\sigma}\frac{1}{2}t_{ij}R_{ij,\alpha}(i-\mathbf{R}_{ij,\alpha}\mathbf{A})\left<c^{\dagger}_{i\sigma}c_{j\sigma}\right>+c.c.
\end{eqnarray}
where $\mathbf{A}$ is the magnetic vector potential and $\alpha=(x,y)$ is the direction of the current. The superfluid density $\rho=\mathbf{J}/\mathbf{A}$ at the limit $\mathbf{A}\rightarrow 0$. Fig.~\ref{cvjk}(c) shows the current as a function of the magnetic vector potential $\mathbf{A}$ at different temperatures. The finite $\mathbf{J}/\mathbf{A}$ ratio is consistent with the Meissner effect, confirming the SC state. Fig.~\ref{cvjk}(d) shows the finite superfluid density $\rho>0$ in the low-temperature region, and $\rho=0$ when $T > T_c$.

In a summary, according to the above experimental quantities, it is evident that the ground state is the gapless nematic SC for $\delta = 0.49$ and $U/W_D = 0.5$.

\begin{figure}[htp]
	\centering
	\includegraphics[width=0.6\textwidth]{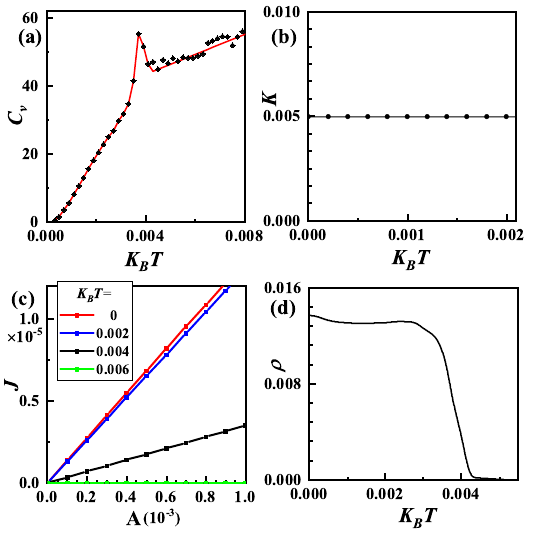}
	\caption{(Color online)  Experiment-relevant quantities for the gapless nematic SC obtained by our calculations. The temperature $K_BT$ is in units of $t$. (a) The specific heat $C_v$ as function of $T$. (b) The NMR Knight-shift $K$ as function of $T$. (c) The current $J$ as function of the exerted vector potential $\mathbf{A}$ at several temperatures. (d) The superfluid density $\rho$ as function of $T$.}\label{cvjk}
\end{figure}

\subsection{The ground state for $\delta = 0.51$ and $U/W_D = 0.35$}
We have confirmed that the ground state is the nematic SC for $\delta=0.49$. For comparison, we have also calculated the ground state properties for another typical filling $\delta=0.51$ and band width $W_D=0.35$. In Fig.~\ref{d51}(a), we show distribution of the amplitude $|\Delta_{mn}|$ ($\Delta_{mn}\in R$ ) of a typical singlet $d_{xy}$-wave pairing gap function between the states $m$ and $n$ (labeled by their energies) near the Fermi level, obtained at the filling $\delta=0.51$. Fig.~\ref{d51}(a) indicates that for each $m$, there is no unique $n$ rendering $|\Delta_{mn}|$ dominates that of any other $n$, violating Anderson's theorem. To determine the realized ground state, we calculate the ground state energy $E$ as a function of the global amplitude $\Delta$ for the $1:r$ (minimized for $r$) and $1:i$ mixing cases. As shown in Fig.~\ref{d51}(b), the energy of the $1:i$ mixing is lower, indicating that the ground state is the chiral SC. Fig.~\ref{d51}(c) shows the local DOS detected by the STM $dI/dV$ curve for a typical site, indicating that both the chiral SC and the nematic SC are gapless.
\begin{figure}[htp]
	\centering
	\includegraphics[width=0.9\textwidth]{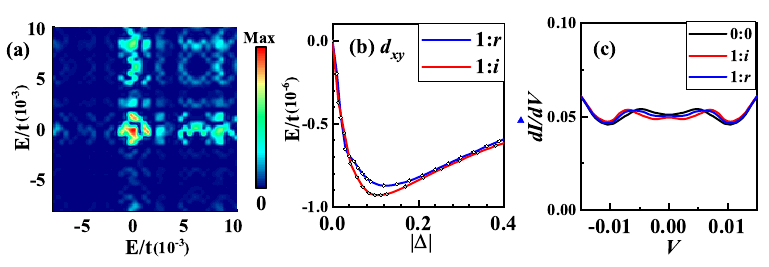}
	\caption{(Color online) Some quantities for $\delta = 0.51$ and $U/W_D = 0.35$. The energy $E$ and $V$ are in unit of $t$. (a) Contour plots of relative $|\Delta_{mn}|$, for a singlet $d_{xy}$-wave state. (b) The SC ground state energy $E$ is a function of the global magnitude $|\Delta|$ of the order parameter, and the mixture of the two degenerate form factors is $1:r$ (bule line) and $1: i$ (red line), where $r$ is a real number. (c) The STM of a typical site for the nematic SC(bule line), chiral SC(red line) and normal state(black line).}\label{d51}
\end{figure}

\section{G-L Theoretical analysis}
The pairing symmetries on the Penrose lattice have been classified according to the irreducible representations (IRRPs) of the $D_5$ point group~\cite{cao}, which includes the 1D $A_1$  (s-wave), $A_2$ (h-wave) and 2D $E_1$ ($(p_x,p_y)$-wave), $E_2$ ($(d_{x^2-y^2}, d_{xy})$-wave) pairings. Here we consider the 2D $E_1$ IRRP, which corresponds to the $(p_x,p_y)$-wave pairing.  The  two basis functions of this pairing are denoted as $(\Delta_{p_{x}},\Delta_{p_{y}})$. For convenience, we rotate the bases and define $\Delta_{\pm}=\Delta_{p_{x}}\pm i\Delta_{p_{y}}$. The general pairing gap function for the $p$-wave is a mixing of $\Delta_{+}$ and $\Delta_{-}$, and should take the form of
\begin{equation}
\Delta=\psi_{+}\Delta_{+}+\psi_{-}\Delta_{-}.
\end{equation}
Fixing the form factor $\Delta_{\pm}$, the free energy $F$ is functional of the global amplitude $\psi_{\pm}$.

The G-L free energy functional $F(\psi_{+}, \psi_{-})$ should be invariant under the rotation $C_5^1$, the U(1) gauge and the mirror-reflection $\sigma$ operations. Under these symmetry operations, the arguments $\psi_{\pm}$ are transformed as
\begin{eqnarray}\label{symmetry}
&&\text{(1) U(1)-gauge :~~~~} \psi_{\pm}\rightarrow e^{i\theta}\psi_{\pm}\nonumber\\
&&\text{(2) C}_5^1\text{-rotation :~~~~} \psi_{\pm}\rightarrow e^{\pm i2\pi/5}\psi_{\pm}\nonumber\\
&&\text{(3)}~~\sigma-\text{mirror :~~~~}\psi_{\pm}\rightarrow \psi_{\mp}.
\end{eqnarray}
The functional $F(\psi_{+}, \psi_{-})$ should be invariant under the above transformations (\ref{symmetry}) on its argument.

Up to $O(\psi_{\pm}^4)$, the form of $F$ allowed by the above symmetries takes the following form,
\begin{eqnarray}
F&=&F^{(2)}+F^{(4)}\nonumber\\
F^{(2)}& = & \alpha(|\psi_{+}|^{2}+|\psi_{-}|^{2}) \nonumber\\
F^{(4)}& = & \beta( |\psi_{+}|^{4}+|\psi_{-}|^{4} ) + \gamma |\psi_{+}|^{2}|\psi_{-}|^{2}
\end{eqnarray}
Consequently, we have
\begin{eqnarray}
F &=& \alpha(|\psi_{+}|^{2}+|\psi_{-}|^{2})+\beta( |\psi_{+}|^{4}+|\psi_{-}|^{4} ) + \gamma |\psi_{+}|^{2}|\psi_{-}|^{2}+o(\psi^6)\nonumber\\
&=& \beta(|\psi_{+}|^2+|\psi_{-}|^2+\alpha/2\beta)^2+(\gamma-2\beta)|\psi_{+}|^2|\psi_{-}|^2+O(\psi^6)
\end{eqnarray}
If $\gamma-2\beta>0$, we get $\psi_{+}=0$ or $\psi_{-}=0$ to minimize the free energy. In this case, the ground state is the chiral SC, such as $p\pm ip$-wave SC. In the contrary, $|\psi_{+}|/|\psi_{-}|=1$ while $\gamma-2\beta<0$, and the ground state is the nematic SC.

To study the effects of the thermal fluctuations around the nematic-SC saddle point, we set $\psi_{+}=e^{i(\theta+\phi)}\psi_{0}$ and $\psi_{-}=e^{i(\theta-\phi)}\psi_{0}$. Here we focus on the low-energy phase fluctuations, and have set the global amplitude $\psi_0>0$ as a constant. The phase fields $\theta$ and $\phi$ are smooth functions of the coarse-grained position $\mathbf{r}$. In order to derive the free energy as an explicit function of $\theta$ and $\phi$, we need to expand the free energy to higher order of the $\psi_{\pm}$ field.

Up to $O(\psi_{\pm}^6)$, the invariance of $F^{(6)}$ under the U(1)-gauge and the $\sigma$-mirror transformations in (\ref{symmetry}) dictates
\begin{eqnarray}
F^{(6)} =  A|\psi_0|^6+B|\psi_0|^4\psi_+\psi_-^*+C|\psi_0|^2\psi^2_+\psi_-^{*2}+D\psi_+^3\psi_-^{*3}+c.c.
\end{eqnarray}
However, under the $C_5^1$ rotation transformation, it is transformed as
\begin{eqnarray}
F^{(6)} \rightarrow  A|\psi_0|^6+Be^{4\pi i/5}|\psi_0|^4\psi_+\psi_-^*+Ce^{8\pi i/5}|\psi_0|^2\psi^2_+\psi_-^{*2}+De^{12\pi i/5}\psi_+^3\psi_-^{*3}+c.c.
\end{eqnarray}
The invariance of $F^{(6)}$ under this transformation dictates $B=C=D=0$. Consequently, $F^{(6)}$ is still not explicit functional of the $\theta$ and $\phi$ fields. The case for $F^{(8)}$ is similar. However, the situation is distinct for $F^{(10)}$, as it can take the following form allowed by the symmetries,
\begin{eqnarray}
F^{(10)} &=& -A_{0}(\psi_{+}^{5}\psi_{-}^{*5}+\psi_{*+}^{5}\psi_{-}^{5})\nonumber\\
&=& -A\cos(10\phi)
\end{eqnarray}
where $A=2A_{0}\psi_{0}^{10}$. Obviously,  $F^{(10)}$ is invariant under all symmetry transformation operations in Eq.(\ref{symmetry}), and it  contributes to the anisotropic part of Hamiltonian in Eq.(5) in the main text.

We can generalize the above derivation to general cases. For the nematic SC on a D$_{2n}$-symmetric lattice ($n\in \mathbf{Z}$), such as on the honeycomb lattice ($n=3$), in order to derive the free energy as an explicit function of the $\theta$ and $\phi$ fields, we need to expand the free energy up to $2n$-th order of its argument $\psi_{\pm}$. The symmetry-allowed $2n$-th order term in the free energy is
\begin{eqnarray}
F^{(2n)}=-A_{0}(\psi_{+}^{n}\psi_{-}^{*n}+\psi_{+}^{*n}\psi_{-}^{n})
\end{eqnarray}
This term contributes to the anisotropy-field part $F^{(2n)}=-2A_{0}\psi_{0}^{2n}\cos(2n\phi)$ in the low-energy classical Hamiltonian. For the nematic SC on a  D$_{2n+1}$ symmetric lattice ($n\in \mathbf{Z}$), such as the Penrose lattice($n=2$). In order to derive the free energy as an explicit function of $\theta$ and $\phi$, we need to expand the free energy up to $2(2n+1)$-th order of its arguments $\psi_{\pm}$, leading to
\begin{eqnarray}
F^{(2(2n+1))}=-A_{0}(\psi_{+}^{2n+1}\psi_{-}^{*2n+1}+\psi_{+}^{*2n+1}\psi_{-}^{2n+1})
\end{eqnarray}
This term contributes to the anisotropy-field part $F^{(2(2n+1))}=-2A_{0}\psi_{0}^{n}\cos[2(2n+1)\phi]$ in the low-energy effective Hamiltonian.

\section{The RG Analysis and More Details}
By the standard RG analysis, the flow equations at the one-loop level are given by:
\begin{eqnarray}
\label{RG-equation}
\frac{dg_{2,0}}{d\ln b}&=&(2-\pi\rho^{'})g_{2,0} \nonumber \\
\frac{dg_{0,2}}{d\ln b}&=&(2-\pi\kappa^{'})g_{0,2} \nonumber \\
\frac{dg_{1,1}}{d\ln b}&=&\left(2-\frac{\pi}{4}(\rho^{'}+\kappa^{'})\right)g_{1,1} \nonumber \\
\frac{dg_{10}}{d\ln b}&=&(2-\frac{25}{\pi\kappa^{'}})g_{10} \nonumber \\
\frac{d\rho^{'}}{d\ln b}&=&-16g_{2,0}^{2}\rho^{'3}-\frac{g_{1,1}^{2}}{2}\rho^{'2}(\rho^{'}+\kappa^{'}) \nonumber \\
\frac{d\kappa^{'}}{d\ln b}&=&\frac{10000g_{10}^{2}}{\pi^4\kappa^{'}}-16g_{0,2}^{2}\kappa^{'3}-\frac{g_{1,1}^{2}}{2}\kappa^{'2}(\rho^{'}+\kappa^{'}).
\end{eqnarray}
Here $b$ is the renormalization scale, $g_{2,0}$, $g_{0,2}$ and $g_{1,1}$ represent the fugacities of the $\theta$-vortices, $\phi$-vortices, and half $\theta$-half $\phi$ vortices. $\rho^{'}=\rho/T$ and $\kappa^{'}=\kappa/T$ represent two kinds of stiffness parametes.

\begin{table}[!h]
\label{tab:1}
\centering
\caption{Fixed points of the coupling parameters under RG, and the corresponding phases. The abbreviations denote: 4e-SC is charge-4e SC; Q-N SC is quasi-nematic SC; Q-N MT is quasi-nematic metal; N-SC is nematic SC; MT is normal metal.}\label{tab:1}
\begin{tabular}{|c|c|c|c|c|c|c|}
  \hline\hline
  $g_{2,0}$ & $g_{0,2}$ & $g_{10}$ & $g_{1,1}$ & $\rho^{'}$ & $\kappa^{'}$ & phase \\
  \hline
  $\infty$ & $\infty$ & 0 & $\infty$ & 0 & 0 & MT\\
  \hline
  0 & $\infty$ & 0 & 0 & finite & 0 & 4e-SC\\
  \hline
  $\infty$ & 0 & 0 & 0 & 0 & finite & Q-N MT\\
  \hline
  0 & 0 & 0 & 0 & finite & finite & Q-N SC\\
  \hline
  0 & 0 & $\infty$ & 0 & finite & $\infty$ & N-SC\\
  %phase & normal & normal & normal & charge-4e SC & quasi-nematic metal & quasi-nematic SC & nematic SC \\
  \hline\hline
\end{tabular}
\end{table}

In Table {\ref{tab:1}}, we present five fixed points of the RG flow Eq.(\ref{RG-equation}) and the corresponding phases, which appear in our numerical results.

\begin{figure}[htp]
	\centering
	\includegraphics[width=0.9\textwidth]{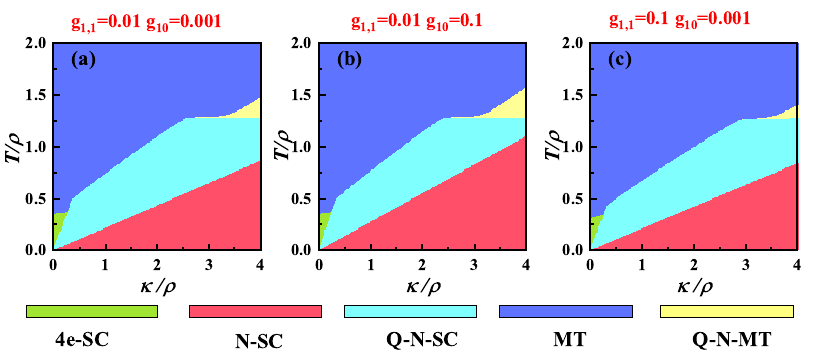}
	\caption{(Color online) The RG phase diagrams presented with different initial coupling parameters. The coupling parameter values are set as $g_{2,0}=g_{0,2}=0.1$, $g_{1,1}=0.01$ and $g_{10}=0.001$ in (a), $g_{2,0}=g_{0,2}=0.1$, $g_{1,1}=0.01$ and $g_{10}=0.1$ in (b), and $g_{2,0}=g_{0,2}=0.1$, $g_{1,1}=0.1$ and $g_{10}=0.001$ in (c).}\label{phase_diag}
\end{figure}
We present more results provided by RG method in Fig.(\ref{phase_diag}) to compare the phase diagrams with different initial values of the coupling parameters. As shown in Fig.(\ref{phase_diag}), we find the regime of the nematic SC and the quasi-nematic metal phase are slightly enlarged with larger anisotropic parameter $g_{10}$. Furthermore, the transition line between the quasi-nematic-SC and the normal metal phase is slightly enhanced while the regime of the quasi-nematic metal phase is slightly suppressed if we increase the fugacities of the half $\phi$-half $\theta$ vortices coupling parameter $g_{1,1}$. On the whole, the topology of the phase diagram is insensitive to the initial values of the coupling parameters.

\section{More Details and Results About the MC Study}

To perform the MC study, we discretize the Hamiltonian Eq.(5) in the main text on a square lattice as
\begin{eqnarray}\label{Hamiltonian_d}
H &=&-\alpha\sum_{\langle ij\rangle}\cos[\theta_{\text{+}}(\mathbf{r}_i)+\theta_{\text{-}}(\mathbf{r}_i)-\theta_{\text{+}}(\mathbf{r}_j)-\theta_{\text{-}}(\mathbf{r}_j)]	 \nonumber\\
&-&\lambda\sum_{\langle ij\rangle}\cos[\theta_{\text{+}}(\mathbf{r}_i)-\theta_{\text{-}}(\mathbf{r}_i)-\theta_{\text{+}}(\mathbf{r}_j)+\theta_{\text{-}}(\mathbf{r}_j)] \nonumber\\
&-&\gamma\sum_{\langle ij\rangle}\cos[\theta_{\text{+}}(\mathbf{r}_i)-\theta_{\text{+}}(\mathbf{r}_j)]+\cos[\theta_{\text{-}}(\mathbf{r}_i)-\theta_{\text{-}}(\mathbf{r}_j)] \nonumber\\
&+& A\sum_{i}\cos[2\theta_{\text{+}}(\mathbf{r}_i)-2\theta_{\text{-}}(\mathbf{r}_i)].
\end{eqnarray}
Here $\langle ij\rangle$ represents nearest-neighbor bonding, and the positive coefficients $\alpha$, $\lambda$ and $\gamma$ satisfy,
\begin{equation}\label{relation}
\alpha=\frac{\rho-2\gamma}{4},~~~~~~~~ \lambda=\frac{\kappa-2\gamma}{4},
\end{equation}
 which ensure that the discretized Hamiltonian (\ref{Hamiltonian_d}) is consistent with the continuous Hamiltonian in the thermodynamic limit. Note that the $\gamma$-term energetically realizes the ``kinematics constraint'' of the $\theta$ and $\phi$ fields on the discrete lattice, which was first proposed in Ref~\cite{Yu_Bo_Liu2023}, and is explained in the following.

 The $\theta(\mathbf{r})$ and $\phi(\mathbf{r})$ fields are related to the $\theta_{\pm}(\mathbf{r})$ fields via the relation $\theta_{\pm}(\mathbf{r})=\theta(\mathbf{r})\pm\phi(\mathbf{r})$. In the continuous space, the physical $\theta_{\pm}(\mathbf{r})$ phase fields should host only integer vortices, which dictates that the $\theta$ and $\phi$ fields should host integer or half-integer vortices simultaneously. This is the ``kinematics constraint'' between the $\theta$ and $\phi$ fields.  On the discrete lattice, the $\alpha(\lambda)$ term energetically allows for integer or half-integer $\theta(\phi)$ vortices, otherwise the energy diverges as $O(L)$ which cannot be compensated by the entropy. For the same reason, the $\gamma$ term only energetically allows for integer $\theta_{+}$ or $\theta_{-}$ vortices, which dictates that the $\theta$ and $\phi$ fields should host integer or half-integer vortices simultaneously. Therefore, the $\gamma$-term energetically imposes the ``kinematics constraint'' between the $\theta$ and $\phi$ fields, which ensures the correct low-energy ``classical Hilbert space'' in the continuum limit. For thermodynamic limit, even an infinitesimal $\gamma$ can energetically guarantee the ``kinematic constraint''. In the MC calculations, we set $\gamma=\frac{1}{4}\rho\kappa/(\rho+\kappa),~A=0.025\rho$, and slight adjustments of the parameters will not qualitatively change the results, including the topology of the phase diagram.

We can determine the phase diagram based on the decaying behavior of the correlation functions $\eta_{\phi/\theta}$. The Table~\ref{tab:2} provides the decaying behavior of the correlation functions $\eta_{\phi/\theta}$ for all possible phases. In the main text, we present the $\eta_{\phi/\theta}$ for the representative B(Q-N SC) and C(Q-N MT) points marked in the MC phase diagram, and their decaying behaviors are consistent with the Table~\ref{tab:2}. As supplements, Fig.~\ref{DE}(a) and (b) show $\Delta r$-dependence of $\eta_{\theta}$ and $\eta_{\phi}$ for the representative point A marked in the MC phase diagram in the main text. Obviously, $\eta_{\phi}$ decays exponentially with $\Delta r$, $\eta_{\theta}$ decays in power law with $\Delta r$, consistent with the properties of the 4e-SC phase. Fig.~\ref{DE}(c) and (d) show the cases for the representative point D marked in the MC phase diagram in the main text: $\eta_{\theta}$ decays in power law with $\Delta r$, $\eta_{\phi}$ saturates to a nonzero value when $\Delta r\to \infty$, consistent with the properties of the N-SC phase. Fig.~\ref{DE}(e) and (f) show the cases for the representative point E marked in the MC phase diagram in the main text. Both correlation functions decay exponentially with $\Delta r$, consistent with the properties of the  MT phase.

\begin{table}[!h]
	\label{tab:2}
	\centering
	\caption{The decaying behavior of the correlation functions $\eta_{\phi}$ and $\eta_\theta$ for all possible phases. The abbreviations denote: 4e-SC is charge-4e SC; Q-N SC is quasi-nematic SC; Q-N MT is quasi-nematic metal; N-SC is nematic SC; MT is normal metal.}\label{tab:2}
	\begin{tabular}{|c|c|c|c|c|c|c|}
		\hline\hline
		Phase & $\eta_\phi$ & $\eta_\theta$ \\
		\hline
		4e-SC & ~~$e^{-r/\xi}$~~ & ~~$r^{-\sigma}$~~ \\
		\hline
		Q-N SC & ~~$r^{-\sigma_1}$~~ & ~~$r^{-\sigma_2}$~~ \\
		\hline
		Q-N MT & ~~$r^{-\sigma}$~~ & ~~$e^{-r/\xi}$~~ \\
		\hline
		N-SC & const & ~~$r^{-\sigma}$~~ \\
		\hline
		MT & ~~$e^{-r/\xi_1}$~~ & ~~$e^{-r/\xi_2}$~~ \\
		\hline\hline
	\end{tabular}
\end{table}

\begin{figure}[h]
	\centering
	\includegraphics[width=0.8\textwidth]{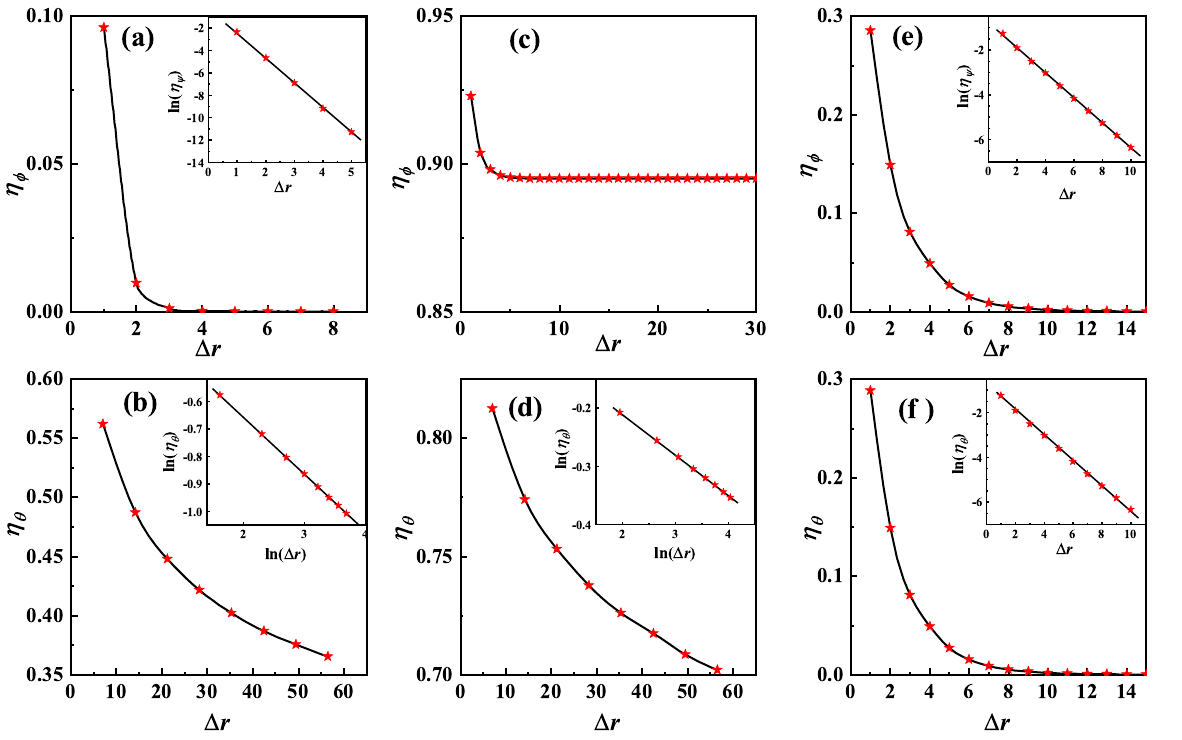}
	\caption{(Color online) The correlation function $\eta_{\phi/\theta}$ for (a) and (b) for A($\kappa=0.2\rho, T=0.2\rho$, representing for the 4e-SC phase), for (c) and (d) for D($\kappa=\rho, T=0.1\rho$, representing for the N-SC phase) and for (e) and (f) for E($\kappa=1\rho, T=0.35\rho$, representing for the MT phase) marked in Fig.2(b) of the main text. Insets of (b,d) the log-log plot, and (a,e-f) only the y- axes are logarithmic. }\label{DE}
\end{figure}

In addition to the correlation functions, some physical quantities can effectively determine the phase diagram and the phase transition temperatures $T_c$. To establish the phase diagram, we calculate the following physical quantities:

1) The specific heat is given as
\begin{eqnarray}\label{Cv2}
C_v=\dfrac{\left\langle H^2\right\rangle -\left\langle H\right\rangle^2}{NT^2}.
\end{eqnarray}
Broad bumps in the specific heat may indicate phase transitions. However, in some cases, the BKT transition is featureless in the $C_v$ curve.

2) The stiffness of the $\theta$-field can be obtained through the approach introduced in Ref.\cite{Zeng2021}. The stiffness $S$ characterizes the superfluid density. Non-zero $S$ indicates the presence of SC.

3) The susceptibility $\chi$ and Binder cumulant $U$ of $\theta$ and $\phi$ are defined as \cite{challa}
\begin{eqnarray}\label{Cv2}
\chi=\dfrac{N(\left\langle m^2\right\rangle-\left\langle m\right\rangle^2)}{K_BT},~~~~ U=1-\dfrac{\left\langle m^4\right\rangle}{3\left\langle m^2\right\rangle^2},
\end{eqnarray}
where $m=\frac{1}{N}\sum_ie^{i\theta}$ for the $\theta$-field or $m=\frac{1}{N}\sum_ie^{i\phi}$ for the $\phi$-field, and $N$ is the lattice-site number. Divergence of $\chi_{\theta/\phi}$ implies $\theta/\phi$ is quasi-long-range order, while finite $\chi_{\theta/\phi}$ indicates $\theta/\phi$ is either long-range order or disorder. The Binder cumulant $U_{\theta/\phi}$ characterizes the order degree of $\theta/\phi$. When the $\theta/\phi$-field is disordered, the quantity $3U_{\theta/\phi}-1=0$; when the $\theta/\phi$-field is long-range ordered or quasi-long-range ordered, the quantity $3U_{\theta/\phi}-1=1$.

In Fig.~\ref{odp}, we show the above quantities as functions of temperature for different lattice sizes at $\kappa/\rho=0.3,1$ and $2.2$. More detailedly, Fig.~\ref{odp}(a1-a3) shows the specific heat $C_v$, Fig.~\ref{odp}(b1-b3) shows the stiffness $S_\theta$, Fig.~\ref{odp}(c1-c3) and (e1-e3) shows the susceptibility $\chi_\theta$ and $\chi_\phi$, and Fig.~\ref{odp}(d1-d3) and (f1-f3) shows the Binder cumulant $3U_\theta-1$ and $3U_\phi-1$, respectively.

For $\kappa/\rho=0.3$, the results are shown in Fig.~\ref{odp}(a1,b1,...,f1). When the temperature $T/\rho$ rises to about 0.05, the specific heat exhibits a finite broad bump, and the susceptibility $\chi_\phi$ changes from finite to divergence, which suggests that the $\phi$-field experiences a BKT phase transition from long-range order to quasi-long-range order at $T/\rho\approx0.05$. The system enters the Q-N SC phase upon this BKT transition. Next, when $T/\rho$ rises to about 0.15, the specific heat exhibits a finite broad bump, the susceptibility $\chi_\phi$ transitions from divergence to finite, and the cumulant $3U_\phi-1$ rapidly drops to zero, which suggesting that $\phi$-field experiences another BKT phase transition from quasi-long-range order to disorder at $T/\rho\approx0.15$. The system enters the 4e-SC phase upon this BKT transition. Finally, when $T/\rho$ rises to about 0.24, the specific heat exhibits a finite broad bump, the stiffness $S_\theta$ rapidly drops to zero, the susceptibility $\chi_\theta$ changes from divergence to finite, and the cumulant $3U_\theta-1$ rapidly drops to zero. These features suggest that the $\theta$-field experiences a BKT phase transition from quasi-long-range order to disorder at $T/\rho\approx0.24$. The system enters the normal MT phase upon this BKT transition.

For $\kappa/\rho=1$, the results are shown in Fig.~\ref{odp}(a2,b2,...,f2). When the temperature $T/\rho$ rises to about 0.143, the specific heat is very smooth, and the susceptibility $\chi_\phi$ changes from finite to divergence, which suggests that the $\phi$-field experiences a BKT phase transition from long-range order to quasi-long-range order at $T/\rho\approx0.143$. The system enters the Q-N SC phase upon this BKT transition. Next, when $T/\rho$ rises to about 0.32, the specific heat exhibits a finite broad bump, the stiffness $S_\theta$ rapidly drops to zero, the susceptibility $\chi_\theta$ and $\chi_\phi$ changes from divergence to finite, and the Binder cumulant $3U_\theta-1$ and $3U_\phi-1$ rapidly drop to zero, which suggests that the $\phi$- and $\theta$- fields simultaneously experience a BKT phase transition from quasi-long-range order to disorder at $T/\rho\approx0.32$. The system enters the normal MT phase upon this BKT transition.

For $\kappa/\rho=2.2$, the results are shown in Fig.~\ref{odp}(a3,b3,...,f3). When temperature $T/\rho$ rises to about 0.31, the specific heat is very smooth, and the susceptibility $\chi_\phi$ changes from finite to divergence, which suggests that the $\phi$-field experiences a BKT phase transition from long-range order to quasi-long-range order at $T/\rho\approx0.31$. The system enters the Q-N SC phase upon this BKT transition. Next, when $T/\rho$ rises to about 0.44, the specific heat exhibits a finite broad bump, the stiffness $S_\theta$ rapidly drops to zero, the susceptibility $\chi_\theta$ changes from divergence to finite, and the Binder cumulant $3U_\theta-1$ rapidly drops to zero. These results suggest that the $\theta$-field experiences a BKT phase transition from quasi-long-range order to disorder at $T/\rho\approx0.44$. The system enters the Q-N MT phase upon this BKT transition.  Finally, when $T/\rho$ rises to about 0.53, the specific heat exhibits a shoulder, the susceptibility $\chi_\phi$ changes from divergence to finite, and the Binder cumulant $3U_\phi-1$ rapidly drops to 0, which suggests that the $\phi$-field experiences a BKT phase transition from quasi-long-range order to disorder at $T/\rho\approx0.53$. The system enters the normal MT phase upon this BKT transition.

\begin{figure}[h]
	\centering
	\includegraphics[width=0.8\textwidth]{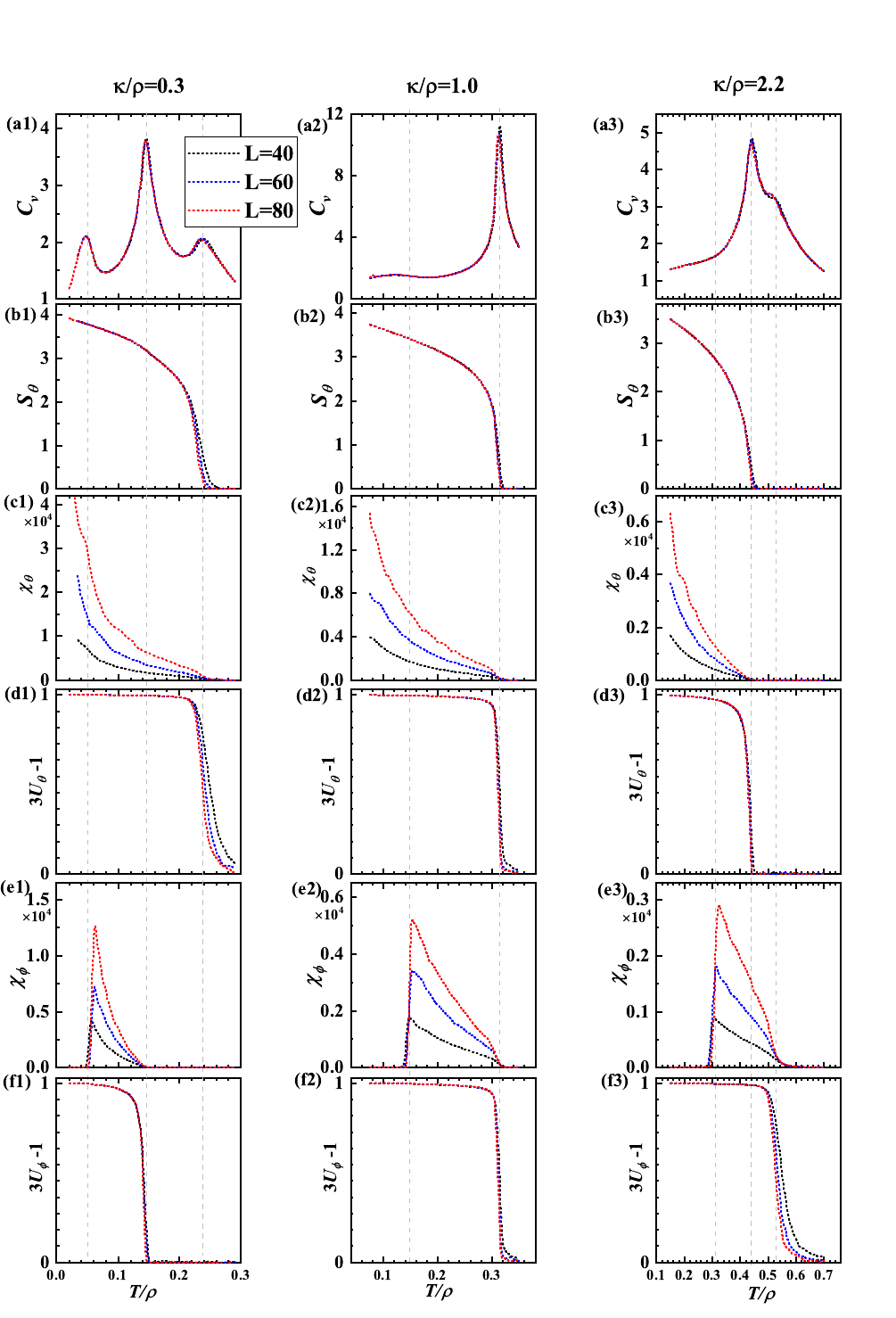}
	\caption{(Color online)Various $T$-dependent quantities for $\kappa/\rho=0.3$ (a1,b1,...,f1), $\kappa/\rho=1$ (a2,b2,...,f2) and $\kappa/\rho=2.2$ (a3,b3,...,f3). The scaling in all figures is $L=$ 40(black line), 60(bule line), and 80(red line). (a1-a3) The specific heat $C_v$. (b1-b3) The stiffness $S_\theta$ of $\theta$. (c1-c3) The susceptibilities $\chi_{\theta}$ of $\theta$. (d1-d3) $3U_{\theta}-1$, where $U_{\theta}$ is the Binder cumulant of the $\theta$-field. (e1-e3) The susceptibilities $\chi_{\phi}$ of $\phi$. (f1-f3) $3U_{\phi}-1$, where $U_{\phi}$ is the Binder cumulant  of the $\phi$-field.}\label{odp}
\end{figure}

\end{widetext}

\end{document}